%% file: main.tex
\documentclass{article}
\usepackage[english]{babel}
\usepackage[utf8]{inputenc}

\usepackage{listings}
\usepackage{color}
 
\definecolor{codegreen}{rgb}{0,0.6,0}
\definecolor{codegray}{rgb}{0.5,0.5,0.5}
\definecolor{codepurple}{rgb}{0.58,0,0.82}
\definecolor{backcolour}{rgb}{0.95,0.95,0.92}
 
\lstdefinestyle{mystyle}{
    backgroundcolor=\color{backcolour},   
    commentstyle=\color{codegreen},
    keywordstyle=\color{magenta},
    numberstyle=\tiny\color{codegray},
    stringstyle=\color{codepurple},
    basicstyle=\footnotesize,
    breakatwhitespace=false,
    breaklines=true,
    captionpos=b,
    keepspaces=true,
    numbersep=5pt,
    showspaces=false,
    showstringspaces=false,
    showtabs=false,
    tabsize=2,
    frame=single
}
\usepackage[font=scriptsize,labelfont=bf]{caption}
\usepackage{float}
\usepackage{graphicx}
\usepackage{hyperref}
\usepackage{indentfirst}
\usepackage{authblk}

\usepackage[backend=bibtex]{biblatex}
\usepackage{geometry}
\title{Replace or Retrieve Keywords In Documents At Scale}

\author{
  Vikash Singh\vspace{-1ex}\\
  \texttt{\href{https://belong.co/}{Belong.co}}\\
  Bangalore, India\\
  \href{mailto:vikash@belong.co}{vikash@belong.co} 
}
\date{}

\begin{document}

\maketitle

\input{abstract}
\noindent\textbf{Subjects} Data Structures and Algorithms (cs.DS)\hfill\break
\newline
\noindent\textbf{Keywords} Information retrieval, Keyword Search, Regex, Keyword Replace, FlashText\hfill\break
\input{introduction}
\input{flashtext_section}
\input{benchmark}
\input{conclusion}
%\input{references}
%\printbibliography

\nocite{*}
\printbibliography
\end{document}

%% file: abstract.tex
\begin{abstract}
In this paper we introduce, the FlashText~\footnote{ FlashText tool for finding or replacing keywords: \url{https://github.com/vi3k6i5/flashtext}} algorithm for replacing keywords or finding keywords in a given text. FlashText can search or replace keywords in one pass over a document. The time complexity of this algorithm is not dependent on the number of terms being searched or replaced. For a document of size \textbf{N} (characters) and a dictionary of \textbf{M} keywords, the time complexity will be \textbf{O(N)}. This algorithm is much faster than Regex (see Figure 1 \& 2), because regex time complexity is \textbf{O(M * N)}. It is also different from Aho Corasick Algorithm~\footnote{ Aho Corasick algorithm: \url{https://en.wikipedia.org/wiki/Aho-Corasick_algorithm}} as it doesn’t match sub-strings.\hfill\break

FlashText is designed to only match complete words (words with boundary characters~\footnote{ ‘\textbackslash b’ represents word boundaries in regex: \url{https://www.regular-expressions.info/wordboundaries.html}} on both sides). For an input dictionary of \textit{\{Apple\}}, this algorithm won’t match it to \textit{‘I like Pineapple’}. This algorithm is also designed to go for the longest match first. For an input dictionary \textit{\{Machine, Learning, Machine learning\}} on a string \textit{‘I like Machine learning’}, it will only consider the longest match, which is \textit{Machine Learning}.\hfill\break

We have made python implementation of this algorithm available as open-source on GitHub, released under the permissive MIT License.\hfill\break

\end{abstract}

%% file: introduction.tex
\section{INTRODUCTION}

In the field of Information Retrieval, keyword search and replace is a standard problem. Often we want to either find specific keywords in text, or replace keywords with standardized names.

\noindent For example:

\begin{enumerate}
 \item Keyword Search: Say we have a resume \textit{(D)} of a software engineer, and we have a list of 20K programing skills \textit{corpus = \{java, python, javascript, machine learning, ...\}}. We want to find which of the 20K terms are present in the resume (\textit{corpus $\cap$ D}). 
 
 \item Keyword Replace: Another common use case is when we have a corpus of synonyms (different spellings meaning the same thing) like \textit{corpus = \{ javascript: [‘javascript’, ‘javascripting’, ‘java script’], ...\}} and we want to replace the synonyms with standardized names in the candidate resume.
  
\end{enumerate}

\begin{figure}[H]
    \centering
    \includegraphics[width=0.6\textwidth]{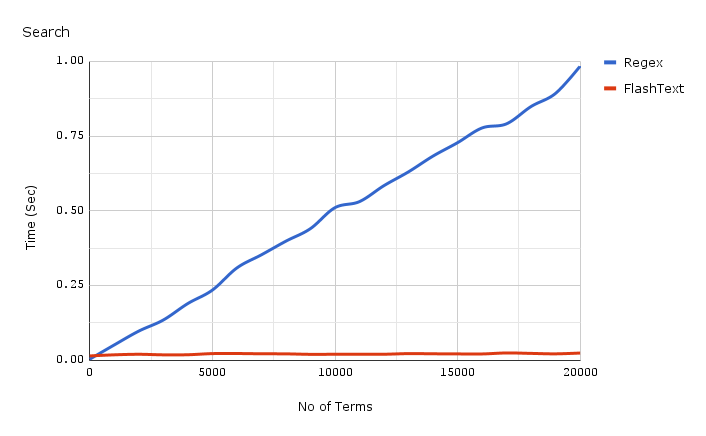}
    \caption{Comparing Time taken (y-axis) by Regex and FlashText to find number of terms (x-axis).}
    \label{fig:search}
\end{figure}

To solve such problems, Regular-Expressions (Regex) are most commonly used. While Regex works well when the number of terms are in 100s, it is, however, considerably slow for 1000s of terms. When the no. of terms are in 10s of thousands and no. of documents are in millions, the run-time will reach a few days. As shown in figure \ref{fig:search}, time taken by Regex to find 15K keywords in 1 document of 10K terms is almost 0.165 seconds. Whereas, for FlashText it is 0.002 seconds. Thus FlashText is 82x faster than Regex for 15K terms.\hfill\break

As the number of terms increase, time taken by Regex grows almost linearly. Whereas time taken by FlashText is almost a constant. In this paper we will discuss the Regex based approach for both keyword search and replace and compare it with FlashText. We will also go through the detailed FlashText algorithm and how it works, and share the code that we used to benchmark FlashText with Regex (which was used to generate the figures in this paper).\hfill\break

\subsection{Regex for keyword search}

Regex as a tool is very versatile and useful for pattern matching. We can search for patterns like '\textbackslash d\{4\}' (which will match any 4 digit number), or keywords like '2017' in a text document. Sample python code (Code \ref{code:ser}) to search '2017' or any 4 digit number in a given string.
\newline
\begin{lstlisting}[language=Python, caption=Sample python code to search 2017 or any 4 digit number in a given string using Regex., label=code:ser]
import re
compiled_regex = re.compile(r'\b2017\b|\b\d{4}\b')
compiled_regex.findall('In 2017 2311 is my birthday.')
['2017', '2311']
\end{lstlisting}

Here '\textbackslash b' is used to denote word-boundary, and is used so that 23114 won’t return 2311 as a match. Word-boundary in Regex ('\textbackslash b') matches special characters like \textit{‘space’, ‘period’, ‘new line’, etc.. \{‘ ‘, ‘.’, ‘\textbackslash n’\}}.

\subsection{Regex for keyword replacement}

We can also use Regex tool to replace the matched term with a standardised term. Sample python code (Code \ref{code:rep}) to replace java script with javascript.

\begin{lstlisting}[language=Python, caption=Sample python code to replace java script with javascript using Regex., label=code:rep]
import re
re.sub(r"\bjava script\b", 'javascript', 'java script is awesome.')
'javascript is awesome.'
\end{lstlisting}

\begin{figure}[H]
    \centering
    \includegraphics[width=0.6\textwidth]{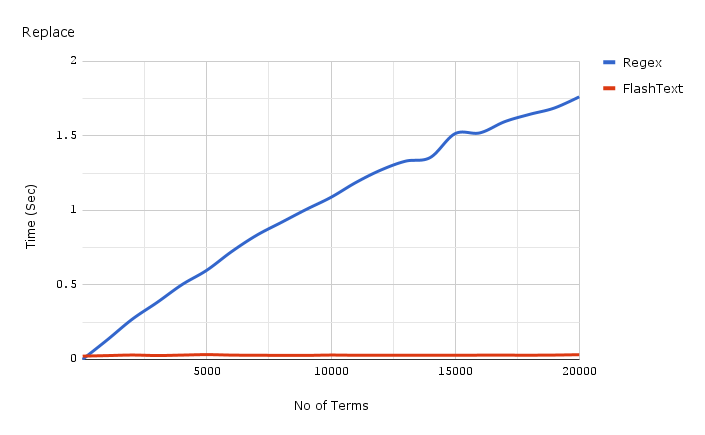}
    \caption{Comparing time taken (y-axis) by Regex and FlashText to replace number of terms (x-axis).}
    \label{fig:replace}
\end{figure}

%% file: flashtext_section.tex
\section{FLASHTEXT}
\begin{figure}[H]
    \centering
    \includegraphics[width=0.8\textwidth]{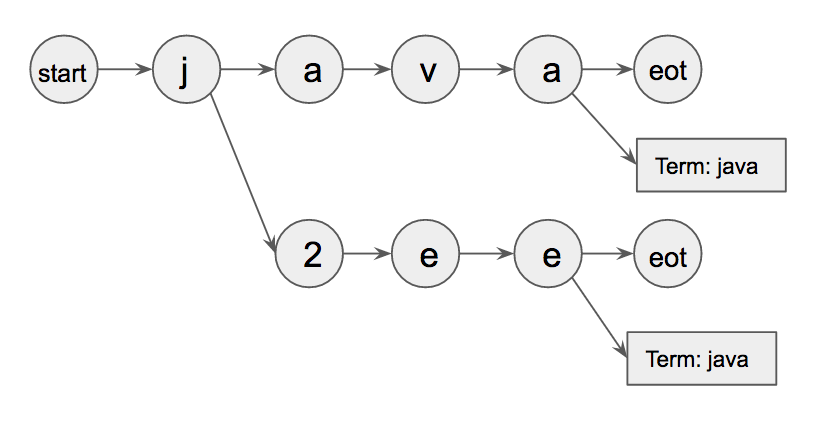}
    \caption{Trie dictionary with 2 keywords, j2ee and java both mapped to standardised term java.}
    \label{fig:trie}
\end{figure}

FlashText is an algorithm based on Trie dictionary data structure and inspired by the Aho Corasick Algorithm. The way it works is, first it takes all relevant keywords as input. Using these keywords a trie dictionary is built (As shown in Figure \ref{fig:trie}).

\textbf{Start} and \textbf{eot} are both special symbols that define word boundary, as defined in Regex. This trie dictionary is used for searching keywords in string as well as replacing keywords in string.

\subsection{Search with FlashText}

For an input string (document), we iterate over it character by character. When a sequence of characters in the document \textbf{\textit{$\textless$\textbackslash b$\textgreater$word$\textless$\textbackslash b$\textgreater$}} match in the trie dictionary from \textbf{\textit{$\textless$start$\textgreater$word$\textless$eot$\textgreater$}} (\textbf{\textit{Start}} and \textbf{\textit{eot}} both stand for word-boundary), we consider it as a complete match. We add the standardized term corresponding to the matched term into a list of keywords found.

\begin{figure}[H]
    \centering
    \includegraphics[width=0.8\textwidth]{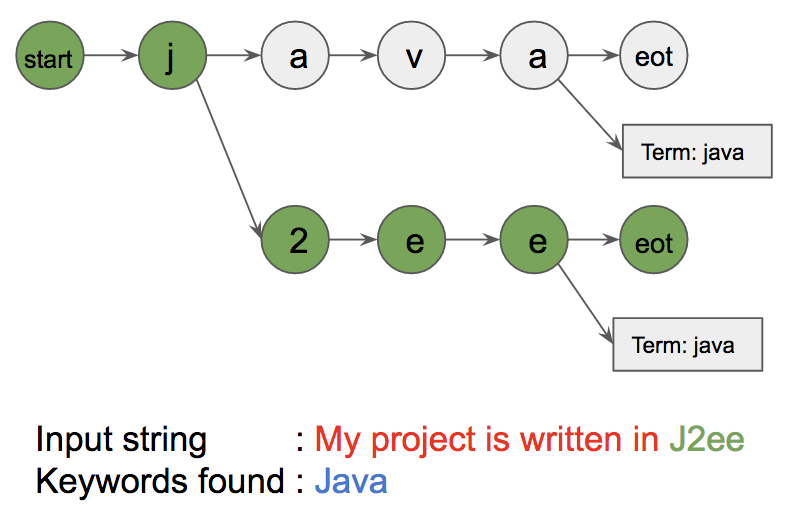}
    \caption{For input string matched character sequence is shown in Green and unmatched in Red.}
    \label{fig:trie_search}
\end{figure}

\subsection{Replace with FlashText}

For an input string (document), we iterate over it character by character. We create an empty return string and when a sequence of characters in the document \textbf{\textit{$\textless$\textbackslash b$\textgreater$word$\textless$\textbackslash b$\textgreater$}} doesn’t match in the trie dictionary, we copy the original word as it is into the return string. When we do have a match, we add the standardised term instead. Thus the return string is a copy of input string, with only matched terms replaced.

\begin{figure}[H]
    \centering
    \includegraphics[width=0.8\textwidth]{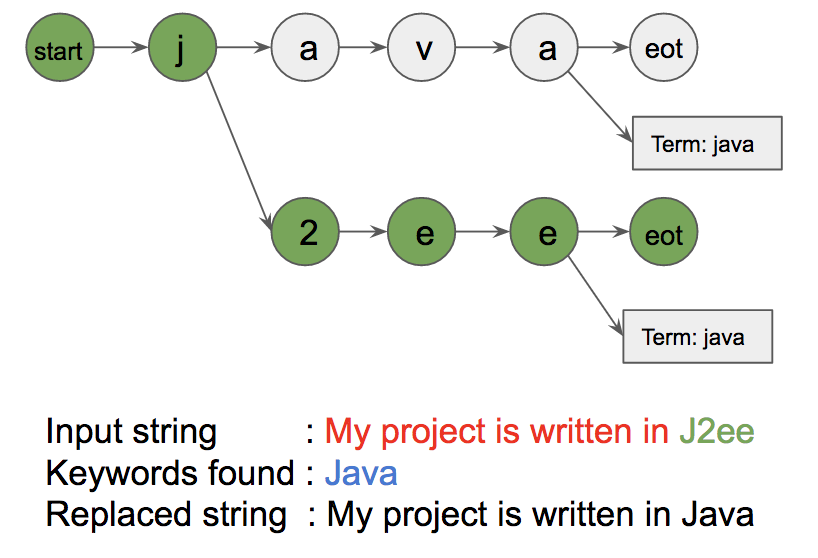}
    \caption{For input string matched character sequence is replaced with standardised name.}
    \label{fig:trie_replace}
\end{figure}

\subsection{FlashText algorithm}
\noindent FlashText algorithm has 3 major parts. We will go over each part separately.
\begin{enumerate}
\item Building the trie dictionary
\item Searching keywords
\item Replacing keywords
\end{enumerate}

\subsubsection{Building the trie dictionary}
To build the trie dictionary, we start with the \textbf{\textit{root}} node which points to an \textit{empty\_dictionary~\footnote{ \texttt{Associative\_array  dictionary data structure: \url{https://en.wikipedia.org/wiki/Associative_array}}}}. This node is used as the start point for all words. We insert a word in the dictionary by inserting the first character to the root node and pointing that to an empty dictionary. The next character from the word, goes as a key in this dictionary, and that again points to an empty dictionary. This process is repeated till we reach the last character in the word. If any character is already present in the dictionary we move to the child dictionary and the next character in the word.
When we reach the end of the word  we insert a special key \textit{\_keyword\_}, to signify end of term (\textbf{\textit{eot}}), and standardized name is stored against this key.

\paragraph*{Input}\hfill\break

\noindent Keyword \textit{w = $c_1$ $c_2$ $c_3$ ... $c_n$} where each \textit{$c_i$} is an input character and \textit{w} is the input keyword.\hfill\break
Standardized name \textit{s} for keyword \textit{w}.
\newpage
\paragraph*{Method}\hfill\break
\begin{lstlisting}[language=Python, caption=Python code for FlashText Initialization and adding keywords to dictionary., label=code:flashinit]
class FlashText(object):
 
    def __init__(self, case_sensitive=False):
        self._keyword = '_keyword_' # end of term (eot) and key to store standardized name
        self._white_space_chars = set(['.', '\t', '\n', '\a', ' ', ','])
        self.keyword_trie_dict = dict()
        self.case_sensitive = case_sensitive
 
    def add_keyword(self, keyword, clean_name=None):
        if not clean_name and keyword:
            clean_name = keyword
 
        if keyword and clean_name:
            # if both keyword and clean_name are not empty.
            if not self.case_sensitive:
                # if not case_sensitive then lowercase the keyword
                keyword = keyword.lower()
            current_dict = self.keyword_trie_dict
            for letter in keyword:
                current_dict = current_dict.setdefault(letter, {})
            current_dict[self._keyword] = clean_name
\end{lstlisting}

\paragraph*{Output}
A dictionary will be created which will look like Figure \ref{fig:trie}.

\subsubsection{Searching for keywords}
Once all keywords are added to the trie dictionary, we can find keywords present in an input string.

\paragraph*{Input}\hfill\break

\noindent String \textit{x = $a_1$ $a_2$ $a_3$ ... $a_n$} where each \textit{$a_i$} is an input character and \textit{x} is the input string.

\paragraph*{Method}\hfill\break
\begin{lstlisting}[language=Python, caption=Python code to get keywords in input string which are present in dictionary., label=code:flashsearch]
def extract_keywords(self, sentence):
    keywords_extracted = []
    if not self.case_sensitive:
        # if not case_sensitive then lowercase the sentence
        sentence = sentence.lower()
    current_dict = self.keyword_trie_dict
    sequence_end_pos = 0
    idx = 0
    sentence_len = len(sentence)
    while idx < sentence_len:
        char = sentence[idx]
        # when we reach a character that might denote word end
        if char not in self.non_word_boundaries:
            # if eot is present in current_dict
            if self._keyword in current_dict or char in current_dict:
                # update longest sequence found
                sequence_found = None
                longest_sequence_found = None
                is_longer_seq_found = False
                if self._keyword in current_dict:
                    sequence_found = current_dict[self._keyword]
                    longest_sequence_found = current_dict[self._keyword]
                    sequence_end_pos = idx
                # re look for longest_sequence from this position
                if char in current_dict:
                    current_dict_continued = current_dict[char]
                    idy = idx + 1
                    while idy < sentence_len:
                        inner_char = sentence[idy]
                        if inner_char not in self.non_word_boundaries and \
                            self._keyword in current_dict_continued:
 
                            # update longest sequence found
                            longest_sequence_found = current_dict_continued[self._keyword]
                            sequence_end_pos = idy
                            is_longer_seq_found = True
                        if inner_char in current_dict_continued:
                            current_dict_continued = current_dict_continued[inner_char]
                        else:
                            break
                        idy += 1
                    else:
                        # end of sentence reached.
                        if self._keyword in current_dict_continued:
                            # update longest sequence found
                            longest_sequence_found = current_dict_continued[self._keyword]
                            sequence_end_pos = idy
                            is_longer_seq_found = True
                    if is_longer_seq_found:
                        idx = sequence_end_pos
                current_dict = self.keyword_trie_dict
                if longest_sequence_found:
                    keywords_extracted.append(longest_sequence_found)
            else:
                # we reset current_dict
                current_dict = self.keyword_trie_dict
        elif char in current_dict:
            # char is present in current dictionary position
            current_dict = current_dict[char]
        else:
            # we reset current_dict
            current_dict = self.keyword_trie_dict
            # skip to end of word
            idy = idx + 1
            while idy < sentence_len:
                char = sentence[idy]
                if char not in self.non_word_boundaries:
                    break
                idy += 1
            idx = idy
        # if we are end of sentence and have a sequence discovered
        if idx + 1 >= sentence_len:
            if self._keyword in current_dict:
                sequence_found = current_dict[self._keyword]
                keywords_extracted.append(sequence_found)
        idx += 1
    return keywords_extracted
\end{lstlisting}

\paragraph*{Output}
A list of standardized names found in the string x, as shown in Figure \ref{fig:trie_search}.

\subsubsection{Replacing keywords}
We can use the same trie dictionary to replace keywords present in an input string with standardized names.

\paragraph*{Input}\hfill\break

\noindent String \textit{x = $a_1$ $a_2$ $a_3$ ... $a_n$} where each \textit{$a_i$} is an input character and \textit{x} is the input string.

\paragraph*{Method}\hfill\break
\begin{lstlisting}[language=Python, caption=Python code for replacing keywords with standardized names\, from dictionary in input string., label=code:flashreplace]
def replace_keywords(self, sentence):
    new_sentence = ''
    orig_sentence = sentence
    if not self.case_sensitive:
        sentence = sentence.lower()
    current_word = ''
    current_dict = self.keyword_trie_dict
    current_white_space = ''
    sequence_end_pos = 0
    idx = 0
    sentence_len = len(sentence)
    while idx < sentence_len:
        char = sentence[idx]
        current_word += orig_sentence[idx]
        # when we reach whitespace
        if char not in self.non_word_boundaries:
            current_white_space = char
            # if end is present in current_dict
            if self._keyword in current_dict or char in current_dict:
                # update longest sequence found
                sequence_found = None
                longest_sequence_found = None
                is_longer_seq_found = False
                if self._keyword in current_dict:
                    sequence_found = current_dict[self._keyword]
                    longest_sequence_found = current_dict[self._keyword]
                    sequence_end_pos = idx
 
                # re look for longest_sequence from this position
                if char in current_dict:
                    current_dict_continued = current_dict[char]
                    current_word_continued = current_word
                    idy = idx + 1
                    while idy < sentence_len:
                        inner_char = sentence[idy]
                        current_word_continued += orig_sentence[idy]
                        if inner_char not in self.non_word_boundaries and \
                            self._keyword in current_dict_continued:
 
                            # update longest sequence found
                            current_white_space = inner_char
                            longest_sequence_found = current_dict_continued[self._keyword]
                            sequence_end_pos = idy
                            is_longer_seq_found = True
                        if inner_char in current_dict_continued:
                            current_dict_continued = current_dict_continued[inner_char]
                        else:
                            break
                        idy += 1
                    else:
                        # end of sentence reached.
                        if self._keyword in current_dict_continued:
                            # update longest sequence found
                            current_white_space = ''
                            longest_sequence_found = current_dict_continued[self._keyword]
                            sequence_end_pos = idy
                            is_longer_seq_found = True
                    if is_longer_seq_found:
                        idx = sequence_end_pos
                        current_word = current_word_continued
                current_dict = self.keyword_trie_dict
                if longest_sequence_found:
                    new_sentence += longest_sequence_found + current_white_space
                    current_word = ''
                    current_white_space = ''
                else:
                    new_sentence += current_word
                    current_word = ''
                    current_white_space = ''
            else:
                # we reset current_dict
                current_dict = self.keyword_trie_dict
                new_sentence += current_word
                current_word = ''
                current_white_space = ''
        elif char in current_dict:
            # we can continue from this char
            current_dict = current_dict[char]
        else:
            # we reset current_dict
            current_dict = self.keyword_trie_dict
            # skip to end of word
            idy = idx + 1
            while idy < sentence_len:
                char = sentence[idy]
                current_word += orig_sentence[idy]
                if char not in self.non_word_boundaries:
                    break
                idy += 1
            idx = idy
            new_sentence += current_word
            current_word = ''
            current_white_space = ''
        # if we are end of sentence and have a sequence discovered
        if idx + 1 >= sentence_len:
            if self._keyword in current_dict:
                sequence_found = current_dict[self._keyword]
                new_sentence += sequence_found
        idx += 1
    return new_sentence
\end{lstlisting}

\paragraph*{Output}
A new string with replaced standardized names found in the string x, as shown in Figure \ref{fig:trie_replace}.

%% file: benchmark.tex
\section{Benchmarking FlashText And Regex}
As shown in Figure \ref{fig:search} and \ref{fig:replace}, FlashText is much faster than Regex. Now we will benchmark and compare FlashText and Regex.

\subsection{Searching keywords}

Python code is used to benchmark search keywords feature. First we will generate a corpus of 10K random words of randomly chosen lengths. Then we will choose 1K terms from the list of words and join them to create a document.\hfill\break

We will choose \textit{k} number of terms terms from the corpus, where \textit{k $\in$ \{0, 1000, 2000, .. , 20000\}}. We will search this list of keywords in the document using both Regex and FlashText and time them.\hfill\break

\begin{lstlisting}[language=Python, caption=Python code to benchmark FlashText and Regex keyword search. Github Gist Link ., label=code:benchmarksearch]
from flashtext import FlashText
import random
import string
import re
import time

def get_word_of_length(str_length):
    # generate a random word of given length
    return ''.join(random.choice(string.ascii_lowercase) for _ in range(str_length))

# generate a list of 100K words of randomly chosen size
all_words = [get_word_of_length(random.choice([3, 4, 5, 6, 7, 8])) for i in range(100000)]

print('Count  | FlashText | Regex    ')
print('------------------------------')
for keywords_length in [0, 1000, 5000, 10000, 15000]:
    # chose 1000 terms and create a string to search in.
    all_words_chosen = random.sample(all_words, 1000)
    story = ' '.join(all_words_chosen)

    # get unique keywords from the list of words generated.
    unique_keywords_sublist = list(set(random.sample(all_words, keywords_length)))

    # compile Regex
    compiled_re = re.compile('|'.join([r'\b' + keyword + r'\b' for keyword in unique_keywords_sublist]))

    # add keywords to Flashtext
    keyword_processor = FlashText()
    keyword_processor.add_keywords_from_list(unique_keywords_sublist)

    # time the modules
    start = time.time()
    _ = keyword_processor.extract_keywords(story)
    mid = time.time()
    _ = compiled_re.findall(story)
    end = time.time()
    # print output
    print(str(keywords_length).ljust(6), '|',
          "{0:.5f}".format(mid - start).ljust(9), '|',
          "{0:.5f}".format(end - mid).ljust(9), '|',)
 
# output: Data for Figure 1
\end{lstlisting}

Github Gist Link for Keyword Search Benchmark code~\footnote{ \texttt{Search Keywords: \url{https://gist.github.com/vi3k6i5/604eefd92866d081cfa19f862224e4a0}}}

\subsection{Replace keywords}

Code to benchmark replace keywords feature.\hfill\break

\begin{lstlisting}[language=Python, caption=Python code to benchmark FlashText and Regex keyword replace. ., label=code:benchmarkreplace]
from flashtext import FlashText
import random
import string
import re
import time

def get_word_of_length(str_length):
    # generate a random word of given length
    return ''.join(random.choice(string.ascii_lowercase) for _ in range(str_length))

# generate a list of 100K words of randomly chosen size
all_words = [get_word_of_length(random.choice([3, 4, 5, 6, 7, 8])) for i in range(100000)]

print('Count  | FlashText | Regex    ')
print('------------------------------')
for keywords_length in [0, 1000, 5000, 10000, 15000]:
    # chose 1000 terms and create a string to search in.
    all_words_chosen = random.sample(all_words, 1000)
    story = ' '.join(all_words_chosen)

    # get unique keywords from the list of words generated.
    unique_keywords_sublist = list(set(random.sample(all_words, keywords_length)))

    # add keywords to Flashtext
    keyword_processor = FlashText()
    for keyword in unique_keywords_sublist:
        keyword_processor.add_keyword(keyword, '_keyword_')

    # time the modules
    start = time.time()
    _ = keyword_processor.replace_keywords(story)
    mid = time.time()
    for keyword in unique_keywords_sublist:
        story = re.sub(r'\b' + keyword + r'\b', '_keyword_', story)
    end = time.time()
    # print output
    print(str(keywords_length).ljust(6), '|',
          "{0:.5f}".format(mid - start).ljust(9), '|',
          "{0:.5f}".format(end - mid).ljust(9), '|',)

# output: Data for Figure 2
\end{lstlisting}

Github Gist Link for Replace Keywords Benchmark code~\footnote{ \texttt{Replace Keywords: \url{https://gist.github.com/vi3k6i5/dc3335ee46ab9f650b19885e8ade6c7a}}}

%% file: conclusion.tex
\section*{CONCLUSION}
As we saw, FlashText is fast and well suited for keyword search/replace. It’s much faster than Regex when the keywords are complete. The complexity of the algorithm is linear in length of the searched text. It is specially useful when the number of keywords is large since all keywords can be simultaneously matched in one pass over the input string.